\documentclass[10pt,preprint,a4paper]{aastex}

\usepackage{graphics,epsf}
\usepackage{amsmath}                
\usepackage{amsfonts}               
\usepackage{amssymb}                
\usepackage{epsfig}                 

\def \cm{~\rm{cm}}
\def \s{~\rm{s}}
\def \km{~\rm{km}}

\def \K{~\rm{K}}
\def \g{~\rm{g}}

\def \AU{~\rm{AU}}
\def \erg{~\rm{erg}}

\def \yr{~\rm{yr}}

\def \day{~\rm{day}}

\def \astrobj#1{#1}

\title{Intermediate luminosity optical transients during the grazing envelope evolution (GEE)}

\author{Noam Soker\altaffilmark{1}}

\altaffiltext{3}{Department of Physics, Technion -- Israel
Institute of Technology, Haifa 32000 Israel;
soker@physics.technion.ac.il.}

\begin{document}

\begin{abstract}
By comparing photon diffusion time with gas outflow time, I argue that a large fraction of the energy carried by the jets during the grazing envelope evolution (GEE) might end in radiation, hence leading to an intermediate luminosity optical transient (ILOT). In the GEE a companion orbiting near the outskirts of the larger primary star accretes mass through an accretion disk, and launches jets that efficiently remove the envelope gas from the vicinity of the secondary star. In cases of high mass accretion rates onto the stellar companion the energy carried by the jets surpass the recombination energy from the ejected mass, and when the primary star is a giant this energy surpasses also the gravitational binding energy of the binary system. Some future ILOTs of giant stars might be better explained by the GEE than by merger and common envelope evolution without jets.
\end{abstract}


\section{INTRODUCTION}
\label{sec:intro}

Numerical hydrodynamical simulations of binary stellar systems orbiting inside a common envelope (CE) have made progress over the years (e.g., \citealt{SandquistTaam1998, Lombardi2006, DeMarcoetal2011, Passy2011, Passyetal2012, RickerTaam2012}; see review by \citealt{Ivanovaetal2013b}). Observations of close binary systems indicate that in many cases a fraction of the energy released by the in-spiraling binary system can unbind the envelope (e.g., \citealt{DeMarcoetal2011, NordhausSpiegel2013}). However, the removal of the CE only by the gravitational energy released by the in-spiraling binary system is not free from problems \citep{Soker2013}. Indeed, recent numerical simulations of the CE phase encountered  difficulties in unbinding the entire envelope (CE; e.g., \citealt{DeMarcoetal2011, Passy2011, Passyetal2012, RickerTaam2012, Ohlmannetal2016}).

These difficulties brought the suggestion that in many cases jets launched by the secondary star,
including main sequence stars, facilitate CE ejection (e.g., \citealt{Soker2013, Soker2014}).
Recent results suggest that even low mass MS stars can indeed accrete at very high rates, up to $\approx 0.01-0.1 M_\odot \yr^{-1}$ \citep{Shiberetal2015} and, furthermore, even if the accreted gas has a specific angular momentum that does not allow for the formation of a Keplerian accretion disk \citep{Soker2004}, the accretion belt formed around the MS star might launch jets \citep{SchreierSoker2016}.
An accretion belt is a sub-Keplerian concentration of accreted gas on the surface of the accreting body closer to the equatorial plane than to the poles. Jets might be blown from the polar regions from where much less gas is accreted \citep{SchreierSoker2016}.

The present paper is aiming at MS stars and brown dwarf that launch jets as they experiencing a grazing envelope evolution (GEE) or at the phase of the onset of a common envelope evolution (CEE). The GEE \citep{Soker2015} takes place in cases where the jets launched by the secondary star remove the entire envelope outside the secondary orbit. The secondary star orbits at the outskirts of the giant star. It can spiral-in to a very small final separation from the giant core, or it might end at a large orbital separation. In the GEE the binary system might be viewed as evolving in a state of "just entering a CE phase". Tidal interaction shrinks the orbit while the jets launched by the secondary star remove the envelope. This makes the GEE an alternative to the CEE in cases where jets efficiently remove the giant envelope \citep{SabachSoker2015}.

Both the GEE and the CEE can in principle account for a small final orbital separation. As well, the system might first experience the GEE, and then the CEE.
When the final orbital separation is large the GEE might better fit the evolution. Such might be the case with the post-AGB star \astrobj{BD$+46^\circ442$} that is in a binary system with an orbital period of $140.77~$days, and that was found by \cite{Gorlovaetal2012} to launch jets. Another example might be \astrobj{IRAS~19135+3937}, a post-AGB binary system with an orbital period of 127 days, and for which \cite{Gorlovaetal2015} reported the detection of a collimated outflow blown by its companion.

Here I study the implications of the interaction of the jets with the very outer parts of the giant envelope for the fraction of energy that is channelled to radiation.
This short study addresses two processes that still need much exploration. These are the powering of some intermediate-luminosity optical transients (ILOTs) and the spiraling-in process of stellar binary systems composed of a giant and a compact object, down to a brown dwarf.
The connection of the CEE with ILOT was mentioned before, e.g., with respect to the ILOTs
\astrobj{OGLE-2002-BLG-360} \citep{Tylendaetal2013} and \astrobj{V1309~Sco}. It was suggested that V1309~Sco resulted from the onset of a CEE \citep{Tylendaetal2011, Ivanovaetal2013a, Nandezetal2014, Kaminskietal2015}. These papers, however, did not consider jets that might be launched from the secondary star. \cite{Ivanovaetal2013a} and \cite{Nandezetal2014} attributed the radiation to recombination energy. \cite{Pejchaetal2016} conducted a general study of binary systems losing mass in the equatorial plane, through the second Lagrangian point $L_2$, or during a CEE. They considered the collision of mass ejected in a spiral pattern in the equatorial plane, and found that about 10 per cents of the kinetic energy is thermalized. They also did not consider jets.

As stellar mergers are common (e.g., \citealt{Kochaneketal2014}) and I expect that many of these go through a GEE at least for part of the time, I continue the exploration of the basic properties of the GEE. Here I consider the radiative efficiency of envelope removal during the GEE. Jets have been proposed before to power ILOTs, including the major outbursts of luminous blue variable (LBV) stars \citep{KashiSoker2010, KashiSoker2016, SokerKashi2012}. Here I concentrate on the GEE.

I compare the jet-powering with recombination-powering in section \ref{sec:Energy}, and in section \ref{sec:rediative} I consider the radiation efficiency from the interaction region of the jets with the envelope. I do not claim in this study that all, or even most, ILOTs are powered by jets. I here only study those ILOTs that are powered by jets during the GEE.
I summarize in section \ref{sec:summary}.

\section{JETS VERSUS RECOMBINATION ENERGY}
\label{sec:Energy}

When an ejected envelope recombines, opacity drops and radiation escapes. This is the reason why  I take the view that recombination energy does not add much to the energy budget in removing a common envelope (\citealt{Harpaz1998, SokerHarpaz2003}; but see, e.g., \citealt{Ivanovaetal2015} and \citealt{Nandezetal2015} for an opposite view). However, the radiation can add significantly to the luminosity of the object. The recombination energy from an ejected ionized mass $M_{\rm eje}$ is
$E_{\rm rec} \simeq 3 \times 10^{46} (M_{\rm eje}/M_\odot) \erg$ for a solar composition (see \citealt{Ivanovaetal2013a} for details).

As recombination cannot eject the envelope, another energy source must be presence. It can be the gravitational energy of the in-spiraling binary system as is traditionally used, it can be the thermal energy of the giant star itself if the CE phase lasts for a sufficiently long time \citep{DeMarcoetal2011}, or jets launched by the mass-accreting companion. When the secondary is still in the outer parts of a giant star, as is the case in the GEE with giant primary stars, the gravitational energy of the binary system is small, and we are left with the energy of jets launched by the accreting companion.
The gravitational energy released by a MS star of mass $M_{\rm MS}$ and radius $R_{\rm MS}$ accreting a mass $M_{\rm acc}$ is
\begin{equation}
E_{\rm acc} \simeq \frac {G M_{\rm MS} M_{\rm acc}}{2 R_{\rm MS}} =
1.9 \times 10^{47} \frac {M_{\rm MS}} {1 M_\odot}
\left( \frac{R_{\rm MS}} {1 R_\odot}  \right)^{-1}
 \frac{M_{\rm acc}} {0.1 M_\odot}
\erg.
  \label{eq:Eaccrete}
\end{equation}
The result is that if the accreted mass is larger than about $ 2 \%$ of the ejected mass $M_{\rm eje}$, then the accreted energy surpass recombination energy. The accreted energy is distributed to radiation and outflow (jets). If we consider the energy carried by the jets only, then the accretion rate should be higher. For example, let us consider jets that are ejected with a terminal speed equals to the escape speed from the accreting companion. For companion mass and radius as scaled in equation  (\ref{eq:Eaccrete}) the escape speed is $620 \km \s^{-1}$. Let us consider also that a fraction $\eta$ of the accreted mass is launched in the jets. In that case the accretion rate should be more than about $5 (0.1/\eta) \%$ of the ejected mass to surpass the recombination energy.

Of course, if the more evolved primary star, of mass and radius $M_1$ and $R_1$, respectively, is still close to the MS, hence its radius is small, then the gravitational energy released by the in-spiraling stellar system can be the dominant energy source
\begin{equation}
E_{\rm binary} \simeq
 \frac {M_1}{10 M_{\rm acc}}
\left( \frac{R_1}{ 10 R_{\rm MS}}  \right)^{-1}
E_{\rm acc}.
  \label{eq:Eaccrete}
\end{equation}
In the present paper I concentrate on cases where the primary star is a giant, $R_1 \ga 10 R_\odot$, and so accretion by the secondary star might be the dominate energy source of the radiation. Hence, I do not deal with \astrobj{V838~Mon} nor with \astrobj{V1309~Sco} where the primary is a MS star or just evolved off the MS.

Part of the energy that is released in the accretion process goes to kinetic energy of jets and disk-wind, and part is radiated away. When the secondary star is deep inside the envelope, as in the CEE, the emitted photons are absorbed, and the expanding envelope loses thermal energy to kinetic energy by adiabatic expansion. In that case most of the radiation (but not the kinetic energy) might indeed come from recombination energy. This is not the case in the GEE as is shown next.

\section{RADIATIVE EFFICIENCY}
\label{sec:rediative}

In the GEE the secondary star launches jets that remove the CE gas from the vicinity of the secondary star \citep{Soker2015}. The gas removal in turn reduces the supply of gas to the accreting secondary star. Hence, the GEE operates in a negative jet-feedback mechanism (JFM). I consider here cases where the high mass accretion rate leads to an ILOT event \citep{KashiSoker2010}. This High-Accretion-Powered ILOT (HAPI) model, is therefore regulated by the JFM \citep{KashiSoker2016}.
This implies in turn, that there is gas in the vicinity of the jets' source, as there should be a reservoir for the accretion disk that launches the jets, but the amount of mass is limited by the operation of the JFM.

The photon diffusion time from a region of depth $H \ll R_1$ in the
envelope that has a density $\rho$ and an optical depth of $\tau =
H \rho \kappa$ is
\begin{equation}
t_{\rm diff}=\frac{3 H}{c} \tau = \frac{3 \kappa \rho H^2}{c},
 \label{eq:diffusion}
\end{equation}
where $\kappa$ is the opacity in that region. In the GEE the jets
launched by the secondary star interact with the regions near the
photosphere of the primary star. The following relation holds in the
photosphere (.e.g., \citealt{KippenhahnWeigert1994})
\begin{equation}
\rho_p \kappa_p R_1 = \frac{2}{3}  \frac{\mu m_H}{k T_1}
 \frac{G M_1}{R_1},
\label{eq:pho}
\end{equation}
where $\rho_p$ is the photospheric density, $\kappa_p$ is the
opacity at the photosphere, and $T_1$, $R_1$, and $M_1$ are the
effective temperature, radius, and mass of the primary star.
Equations (\ref{eq:diffusion}) and (\ref{eq:pho}) can be combined
to yield the relation
\begin{equation}
t_{\rm diff-MS} \approx 0.05   
 \left( \frac{H}{0.1R_1} \right)^2
 \left( \frac{\rho}{\rho_p} \right)
 \left( \frac{\kappa}{40 \kappa_p} \right)
 \left( \frac{T_p}{6000 \K} \right)^{-1}
 \left( \frac{M_1}{M_\odot} \right) \day ,
 \label{eq:diffusionMS}
\end{equation}
that was scaled for a solar-like star with a photospheric opacity
of $\kappa_p \approx 0.01 \cm^2 \g^{-1}$. The opacity at outburst
was taken to be $\kappa=0.4 \cm^2 \g^{-1}$. The ratio $H/R_1$ was
taken as that of a low mass MS star or a BD.
We can scale quantities for AGB stars with $\kappa_p \approx
0.001$, and so
\begin{equation}
t_{\rm diff-AGB} \approx 1   
 \left( \frac{H}{0.1R_1} \right)^2
 \left( \frac{\rho}{\rho_p} \right)
 \left( \frac{\kappa}{400 \kappa_p} \right)
 \left( \frac{T_p}{3000 \K} \right)^{-1}
 \left( \frac{M_1}{M_\odot} \right) \day .
 \label{eq:diffusionAGB}
\end{equation}
The ratio $H/R_1 =0.1$ was taken as the ratio of the pressure
scale hight in the outer atmosphere of AGB star to the radius.

Equations (\ref{eq:diffusionMS}) and (\ref{eq:diffusionAGB}) are
very crude. The density outside the photosphere is much lower than
$\rho_p$, and that inward to the photosphere is much higher than
$\rho_p$. After expansion, the density can decrease further.

The outflow time depends on the acceleration of the envelope to a
velocity $v_{\rm out}$. The companion is taken to be a BD or a low mass MS star
in the case of a primary star just off the MS, and a MS star in the case of an
AGB primary star. The jets from the companion accelerate gas along the polar directions and in the lobe they inflate  lobes  to a velocity of $v_{\rm out}$
that is a fraction of the jets speed of $\approx 600 \km \s^{-1}$.
The outflow time is then
\begin{equation}
t_{\rm flow} \approx \frac{R_1}{v_{\rm out}} = 0.016
 \left( \frac{R_1}{1 R_\odot} \right)
 \left( \frac{v_{\rm out}}{500 \km \s^{-1}} \right)^{-1} \day
= 3.5
 \left( \frac{R_1}{1 \AU} \right)
 \left( \frac{v_{\rm out}}{500 \km \s^{-1}} \right)^{-1} \day,
 \label{eq:test}
\end{equation}
where the outflow speed is just a little below the escape speed from the main sequence companion that launches the jets. 

 The conclusion from the above crude estimates is that in the GEE
the jets accelerate the outskirts of the primary envelope such
that in some case the flow time might be of the order of the diffusion time
\begin{equation}
t_{\rm flow} \approx  t_{\rm diff}.
 \label{eq:test}
\end{equation}
The implication is that a non negligible fraction of the kinetic
energy of the jets in the GEE can be transferred to radiation energy. This
is unlike the case in supernovae explosions, where the radiated
energy from the explosion is a tiny fraction of the kinetic eneregy
(unless later there is a collision with circum-stellar matter).

A similar relation holds in the case that the jets are launched
outside the AGB envelope, but they collide with the wind at a
distance of $\approx 10 \AU$ from the binary system
\citep{AkashiSoker2013}.


\section{SUMMARY}
\label{sec:summary}

In this study I continued to explore the properties of the recently proposed GEE \citep{Soker2015}.  In the GEE a companion orbiting near the outskirts of the larger primary star accretes mass through an accretion disk. Jets that are launched by the accretion disk efficiently remove the envelope gas in the vicinity of the secondary star. This takes place via a negative jet-feedback mechanism (JFM). The jets interact with the envelope and heat and accelerate the gas.

In section \ref{sec:rediative} I showed that a large fraction of the energy carried by the jets may very well end in radiation. The radiation might be observed as a transient event called ILOT (other names are Red Novae;  Luminous Red Novae; Red Transients; Intermediate-Luminous Red Transients; see discussion of names in \citealt{KashiSoker2016}).
This type of ILOT events that take place during the GEE, is one that is powered by high mass accretion (\citealt{KashiSoker2010}; the HAPI model), and that is regulated by the JFM. \cite{KashiSoker2016} proposed that some ILOTs, not only those that occur during the GEE, are regulated by the JFM, and compared them with other astrophysical systems where the JFM operates.

Although \cite{KashiSoker2016} mentioned also ILOTs that occur during a GEE, they did not go into the properties of the GEE-powered ILOTs. In the present paper I closed this gap by pointing out that the photon diffusion time during the GEE favors the occurrence of an ILOT event. In section \ref{sec:Energy} I pointed out that in some cases the energy carried by the jets might surpass the recombination anergy from the ejected mass, and when the primary star is a giant this energy surpasses also the gravitational energy of the binary system.
In many of the cases with AGB primary stars a bipolar planetary nebula will be formed \citep{SokerKashi2012}. Relevant to that claim is the suggestion made by \cite{Humphreysetal2011}  that the ILOT NGC~300~2008OT-1 had a bipolar outflow; see however the claim by \cite{Adamsetal2015} that NGC~300~2008OT-1 might have been a SN explosion rather than an ILOT event.

I summarize by calling future studies that will claim for a merger and/or a CEE explanation for newly discovered ILOTs (Red Novae; Luminous Red Novae; Red Transients; Intermediate-Luminous Red Transients), particulary those with giant primary stars, to consider the GEE as a viable alternative.


\end{document}